\documentclass[a4paper,11pt]{article}

\pdfoutput=1 % if your are submitting a pdflatex (i.e. if you have images in pdf, png or jpg format)

\usepackage[utf8]{inputenc}
\usepackage{jinstpub}
\usepackage{pgf}
\usepackage[normalem]{ulem}\RequirePackage{lineno}
\usepackage{color}
\usepackage{upgreek}

\begin{document}

%%%%%%%%%%%%%%%%%%%%%%%%%%%%%%%%%%%
%%% Title Page
%%%%%%%%%%%%%%%%%%%%%%%%%%%%%%%%%%%

% title
\title{A High-Sensitivity Radon Emanation Detector System for Future Low-Background Experiments}

% authors
\author{Daniel Wiebe\footnote{nee: Daniel Baur},}\emailAdd{daniel@baurclan.de}
\author{Sebastian Lindemann,}\emailAdd{sebastian.lindemann@physik.uni-freiburg.de}
\author{Marc Schumann}
\affiliation{Physikalisches Institut, Universit\"at Freiburg, 79104 Freiburg, Germany}

% abstract
\abstract{
Radioactive radon atoms originating from the primordial $^{238}\mathrm{U}$ and $^{232}\mathrm{Th}$ decay chains are constantly emanated from the surfaces of most materials.
The radon atoms and their radioactive daughter isotopes can significantly contribute to the background of low-background experiments. The $^{222}\mathrm{Rn}$ progeny $^{214}\mathrm{Pb}$, for example, dominates the background of current liquid xenon-based direct dark matter detectors.
We report on a new detector system to quantify the $^{222}\mathrm{Rn}$ surface emanation rate of materials.
Using cryogenic physisorption traps, emanated radon atoms are transferred from an independent emanation vessel and concentrated within the dedicated detection vessel. The charged radon daughter isotopes are collected electrostatically on a silicon PIN photodiode  to spectrometrically measure the alpha decays of $^{214}\mathrm{Po}$ and $^{218}\mathrm{Po}$.
%inside the dedicated detection vessel, where the charged daughter isotopes, most importantly $^{214}\mathrm{Po}$ and $^{218}\mathrm{Po}$, are electrostatically collected and detected on a silicon PIN photodiode.
The overall detection efficiency is $\sim$36\,\% for both polonium channels.
%The intrinsic detection vessel background was measured to be $\sim 2.4\,\mathrm{cpd}$ ($28\,\mathrm{\upmu Bq}$) and $\sim 1.5\,\mathrm{cpd}$ ($17\,\mathrm{\upmu Bq}$) for $^{218}\mathrm{Po}$ and $^{214}\mathrm{Po}$, respectively.
The radon emanation activity of the emanation vessel was measured to be $(0.16\pm 0.03)\,\mathrm{mBq}$, resulting in a detection sensitivity of $\sim$0.06\,$\mathrm{mBq}$ at $90\,\%$ C.L..
}

% keywords

\maketitle

%%%%%%%%%%%%%%%%%%%%%%%%%%%%%%%%%%%
%%% Introductions
%%%%%%%%%%%%%%%%%%%%%%%%%%%%%%%%%%%

\section{Introduction} %: Radon-Induced Background in Rare-Event Searches}
\label{sec:introduction}

%One of the leading technologies to directly search for ultra rare processes, such as WIMP dark matter or neutrinoless double beta decay, are time projection chambers (TPCs) filled with cryogenic liquid xenon (LXe). 
Time projection chambers (TPCs) filled with cryogenic liquid xenon (LXe) are one of the leading technologies to directly search for ultra-rare processes, such as neutrinoless double beta decay or WIMP dark matter scattering.
% direct WIMP dark matter detection with dual-phase LXe TPCs
The best constraints on spin-independent WIMP-nucleon scattering for WIMP masses above $5\,\mathrm{GeV}/c^2$ to date come from dual-phase LXe TPCs with multi tonne-scale targets~\cite{xenon1t_18,PandaX-4T:2021bab,lz_collaboration__wimp_results__2022, xenon_collaboration__wimp_results__2023}.
%These instrument tonne-scale liquid xenon (LXe) targets to detect both the scintillation and ionization signals generated by the interaction of particles with the target xenon atoms. The ratio of light to charge signal is used to distinguish between nuclear recoil (NR, WIMP-like) and electronic recoil (ER, background-like) signals.
One of the most critical backgrounds in these experiments arises from the decay of $^{222}\mathrm{Rn}$ (radon) and its progenies. Radon is part of the ubiquitous primordial $^{238}\mathrm{U}$ decay chain and emanates off any detector component via recoil ejection and diffusion. Due to its comparatively long half-life of $3.8\,\mathrm{d}$ and its chemical inertness, it distributes within the LXe target and cannot be mitigated via target fiducialization. For similar reasons, radon contributes to the background in current and future searches for the neutrinoless double beta decay of $^{136}\mathrm{Xe}$~\cite{EXO-200:2014ofj,NEXT:2018zho,nEXO:2021ujk}, where radon emanation and subsequent high-energy decays result in backgrounds in the region of interest around $Q_{\beta\beta}$.
%The background in LXe-based dark matter detectors is from the radon-daughter $^{214}\mathrm{Pb}$, which beta-decays with a $9.3\,\%$ branching ratio directly to the ground state without the emission of a coincident gamma ray from a nuclear de-excitation which could be identified to reject the background event.

%One of the most critical backgrounds in these experiments is the leakage of low-energetic beta decays of $^{214}\mathrm{Pb}$ (ER signals), a daughter of $^{222}\mathrm{Rn}$, into the WIMP NR-signal region:
%Radon is part of the ubiquitous primordial  $^{238}\mathrm{U}$ decay chain and emanates off any detector component via recoil ejection and diffusion.
%Due to its comparably long half-live of $3.8\,\mathrm{d}$ and its chemical inertness, it distributes within the LXe target.
%$9.3\,\%$ of the $^{214}\mathrm{Pb}$ daughter atoms beta-decay directly to the ground state.
%Consequently, these radon-induced ER background events can neither be mitigated via target fiducialization nor by tagging coincident nuclear de-excitations with emission of a gamma ray.
The XENONnT dark matter experiment has recently reported a radon activity concentration of $1.8\,\upmu\mathrm{Bq}/\mathrm{kg}$~\cite{xenoncoll22}, which has been further reduced below $1\,\upmu\mathrm{Bq}/\mathrm{kg}$ in the meanwhile~\cite{xenon_collaboration__wimp_results__2023}.  
% future
Future dark matter experiments with a LXe target above $40\,\mathrm{t}$, such as DARWIN~\cite{darwin_16} or XLZD~\cite{Aalbers:2022dzr}, aim at exploring the entire WIMP parameter space accessible to the LXe TPC technology~\cite{billardea_14, darwin_15} and offer an interesting neutrino physics program~\cite{Baudis:2013qla,DARWIN:2020jme,DARWIN:2020bnc}.
To reach the design sensitivity, their %ER and NR 
background must be dominated by irreducible interactions of solar and atmospheric neutrinos~\cite{darwin_16}.
This requires reducing the concentration of $^{222}\textrm{Rn}$ to $0.1\,\upmu\mathrm{Bq}/\mathrm{kg}$~\cite{darwin_15}, corresponding to an order-of-magnitude improvement compared to the current-generation~\cite{xenon_collaboration__wimp_results__2023}. The planned double beta experiment nEXO aims for a similar $^{222}\textrm{Rn}$ concentration of about $0.25\,\upmu\mathrm{Bq}/\mathrm{kg}$~\cite{nEXO:2021ujk}. 

This challenging goal will be met by a combination of background mitigation methods: surface treatment~\cite{Bruenner:2020arp}, detector design~\cite{Sato:2019qpr,yuehanea_20,Dierle:2022zzh}, active radon removal~\cite{Murra:2022mlr, Abe2012, Aprile2017} as well as by using only ultra-low-emanation materials for all detector parts in direct contact with xenon.
%Radon emanation depends on material properties and primarily probes surface contamination, thus emanation rates must be quantified by means of highly sensitive radon emanation detectors instead of potentially misleading bulk measurements of the $^{226}\textrm{Ra}$ activity via standard gamma spectrometry.
%Radon emanation depends on material properties and primarily probes surface and skin contamination. 
Radon emanation depends on material properties and is often only a measure of surface contamination.
Therefore, emanation rates must be quantified using highly sensitive radon emanation detectors, rather than relying on potentially misleading bulk measurements of $^{226}\textrm{Ra}$ activity via standard gamma spectrometry.

% radon emanation chamber
The concept of the \textit{electrostatic radon emanation chamber} \cite{albrecht_67,Wojcik2017} used in this work is by now an integral part of the radiopurity assay of modern rare-event searches~\cite{XENON:2020fbs,lzcol20, 2020:laubenstein, XENON:2021mrg,  PandaX-4T:2021lbm}.
Such an instrument consists of a gas-tight vacuum vessel that houses a silicon PIN photodiode set to negative high-voltage, creating an electrical drift field with respect to the vessel on ground potential.
$^{222}\mathrm{Rn}$ atoms present in the vessel will eventually decay, leaving a fraction of the daughter $^{218}\mathrm{Po}$ in a positively charged state as a consequence of the alpha decay recoil~\cite{Howard1991,Pagelkopf2003}.
The ionized daughters are then collected electrostatically on the surface of the PIN diode, 
where the isotopes $^{218}\mathrm{Po}$ and $^{214}\mathrm{Po}$ are identified by measuring the induced charge signal proportional to the energy deposited by their alpha decays.
%Therein, the alpha decays of the radon daughters $^{218}\mathrm{Po}$ and $^{214}\mathrm{Po}$ generate a number of electron-hole pairs that is proportional to their characteristic energies, allowing the spectrometric identification of either species.
Modeling the evolution of the detected activity of the radon daughters during the measurement allows the inference of the sample's radon emanation rate $\mathcal{R}$.

% paper outline
This work presents the design, construction, and performance of the \textit{MonXe} radon emanation detector. %, developed for the radiopurity assay for the future DARWIN experiment.
%\textcolor{blue}{Sections~\ref{sec:setup} and \ref{sec:detector_operation_and_model} describe the detector's experimental setup and operation principle, respectively.}
Section~\ref{sec:setup} describes the working principle and experimental setup of the detector. Section~\ref{sec:detector_operation_and_model} explains the operation of the instrument. 
Section~\ref{sec:performance} elaborates on its performance in terms of background, efficiency, and sensitivity.
Section~\ref{sec:sample_screening_measurements} presents exemplary screening measurements of high-activity zeolite granulate and a low-activity PTFE sample.
The article concludes in Section~\ref{sec:conclusio} with a summary and an outlook on future detector optimizations.

%%%%%%%%%%%%%%%%%%%%%%%%%%%%%%%%%%%
%%% Experimental Setup
%%%%%%%%%%%%%%%%%%%%%%%%%%%%%%%%%%%

\section{Experimental Setup}
\label{sec:setup}

% general overview

\begin{figure}
    \centering
    \includegraphics[width=\textwidth]{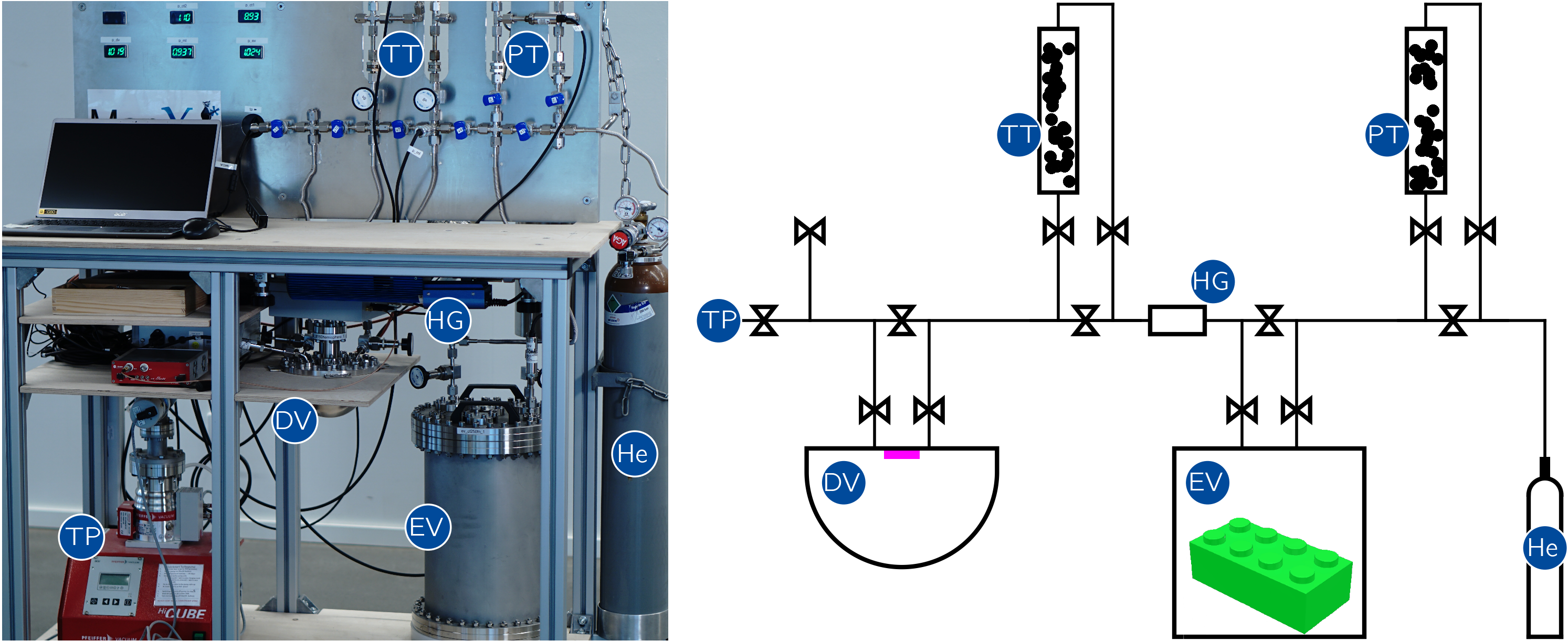}
    \caption{
        Photograph \textbf{(Left)} and schematic \textbf{(Right)} of the MonXe radon emanation detector. The examined sample (green brick) is placed inside the emanation vessel (EV).
        The emanated radon atoms are transferred into the hemispherical detection vessel (DV), where the activity is measured using a PIN photodiode (magenta).
        The transfer occurs by evacuating the emanation vessel using a turbo molecular pump (TP) through the transfer trap (TT), which is cooled to liquid nitrogen temperature. In this trap radon atoms adsorb onto activated charcoal (black spheroids).
        A hot zirconium getter (HG) installed in series removes other contaminants such as O$_2$ or H$_2$O in the gas.
        By heating the transfer trap (TT), the radon atoms desorb again and are guided into the detection vessel by a flow of helium, which is purified by cryogenic activated charcoal housed inside the purification trap (PT).
    }
\label{fig:gas_system_and_vessel}
\end{figure}

\begin{figure}[h!]
    \centering
    \includegraphics[width=\textwidth]{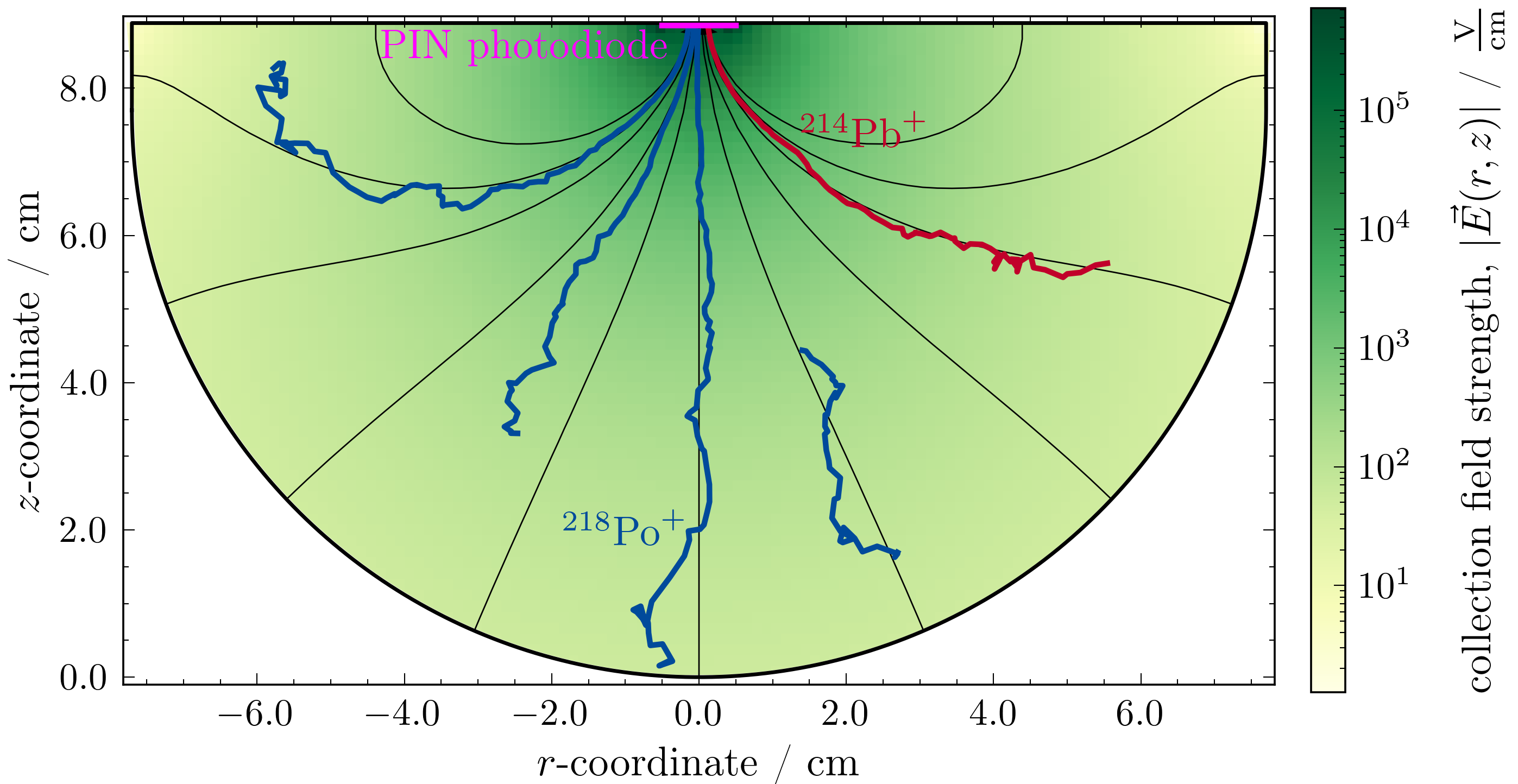}
    \caption{
        Cross-sectional view of the hemispherical MonXe detection vessel (DV), showing the simulated electrostatic drift field (field lines: solid black; field strength: color map) and particle tracks.
        %The simulated electrostatic drift field  is established between the negatively biased silicon PIN photodiode and the vessel on ground. 
        $^{222}\mathrm{Rn}$ atoms in the vessel eventually decay and their
        positively charged daughters $^{218}\mathrm{Po}$ drift towards the photodiode (solid blue tracks).
        If they neutralize along their path, the $^{218}\mathrm{Po}$ atoms propagate via diffusion (also simulated but not shown) until they alpha-decay into positively charged $^{214}\mathrm{Pb}$ and can be collected again (solid red track).
        The simulations suggest a collection efficiency close to $100\,\%$ for either polonium species for a collection voltage of $-1\,\mathrm{kV}$~\cite{wb21}.
    }
\label{fig:drift_sim}
\end{figure}

%The MonXe radon detector system features a set of two decoupled vacuum vessels, where one is used to accommodate the emanating sample and the second one is used for the actual radon detection.
%for the emanation and detection of radon, respectively.
%This allows assessing the emanation of large and bulky samples while optimizing the radon detection in a reproducible measurement procedure. 
%The use of two vessels facilitates the measurement of radon emanation rates from large samples while optimizing radon detection through a consistent and reproducible measurement procedure.
%The use of two vessels facilitates the measurement of radon emanation rates from large samples in one vessel while optimizing the collection efficiency of charged radon progeny in the other vessel.
The MonXe radon detector system comprises two decoupled vacuum vessels: one for accommodating the emanating sample and the other for actual radon detection. This dual-vessel setup facilitates measuring radon emanation rates from large samples in one vessel while optimizing the collection efficiency of charged radon progenies in the other.
A photograph of the detector system is shown in the left panel of Figure~\ref{fig:gas_system_and_vessel}. The gas system connecting both vessels is sketched in the right panel.
The hemispherical shape of the detection vessel (DV), with a radius of $7.7\,\mathrm{cm}$ and a volume of $1.2\,\mathrm{l}$, was optimized in terms of electrostatic collection efficiency via dedicated particle-tracking simulations taking into account diffusion effects, as shown in Figure~\ref{fig:drift_sim}.
The PIN diode is installed in the central bore of the vessel's CF160 flange, with the diode surface being aligned with the inner flange plane.
The CF160 flange features six additional CF16 flanges to connect sensors and the concentration line.
The detection vessel's inner surface is electropolished to minimize its intrinsic radon emanation.
The cylindrical emanation vessel (EV) has a height of $40.0\,\mathrm{cm}$ and an inner diameter of $25.4\,\mathrm{cm}$, corresponding to a volume of $20.4\,\mathrm{l}$. It is closed off with CF250 flanges on both sides.
The entire system was built from CF/Conflat and VCR metal-sealed UHV components and exhibits a leak rate below $10^{-9}\,\mathrm{mbar\,l}/\mathrm{s}$.

% charcoal traps
Samples of arbitrary material, size, and shape can be placed inside the emanation vessel, which is evacuated after installation of the sample, and afterwards refilled with $1\,\mathrm{bara}$ of purified helium.
In principle, several emanation vessels could be installed in parallel to speed up an extensive measurement campaign.
The emanation rate of a sample is assessed by transferring the emanated radon atoms into the detection vessel via cryogenic physisorption on activated charcoal  (charcoal: Bl\"ucher Saratech 100050-VC000021). 
Helium (grade 5.0) is used as carrier gas; the gas bottle is directly attached to the  \textit{purification trap} (PT), which is kept at cryogenic temperature during operation by immersing it into a liquid nitrogen bath.
This purifies the helium gas introduced into the system, removing radon and other contaminants.
The emanation vessel is filled with helium gas which is subsequently extracted from the system by means of a vacuum pump via the cold \textit{transfer trap} (TT), also kept at liquid nitrogen temperature, where the radon atoms are adsorbed.
The transfer (purification) trap is made of an electropolished stainless steel cylinder of 10.0\,cm length and 1.2\,cm (4.0\,cm) inner diameter to accommodate 10\,g (75\,g) of activated charcoal. 
By heating the transfer trap to $175^{\circ}\mathrm{C}$, the radon gets desorbed and is flushed into the detection vessel using purified helium gas until the normal operating pressure of $1\,\mathrm{bara}$ is reached.

% sensors and getter
Pressure sensors (OMEGA PX409) monitor the emanation and detection vessels during the radon transfer and measurement phases.
A hot zirconium getter (SAES MonoTorr PS3-MT3-R-2), installed between the emanation vessel and the gas system's main line, removes impurities outgassing from the sample that could otherwise neutralize the charged radon progeny and reduce the electrostatic collection efficiency.
%Otherwise, these impurities might lead to the neutralization of the charged radon progenies, which in turn would reduce the collection efficiency.

%\subsection{Data Acquisition Chain}
%\label{ssec:daq}

% DV
%The hemispherical geometry of the detection vessel was studied and optimized via electrostatic simulations (Comsol Multiphysics) to yield a collection efficiency of almost $100\,\%$ for either polonium species~\cite{wb21}.
%Figure~\ref{fig:drift_sim} shows the electrical collection field and simulated particle tracks.
% photodiode
The silicon PIN photodiode (Hamamatsu S3590-09 \cite{diode_datasheet}) installed inside the detection vessel has a photosensitive area of $10\,\mathrm{mm}\times 10\,\mathrm{mm}$.
The diode does not feature a protective epoxy cover but directly exposes its p-layer to minimize the absorption of the energy of the impinging alpha particles in an inactive material layer.
% frontend electronics module
The diode is embedded in a PTFE cylinder installed in a CF40 double nipple centered on the flange of the detection vessel. 
%The diode is attached to a solid PTFE cylinder installed in a CF40 double nipple centered on the flange of the detection vessel.
Its surface aligns with the level of the flange's vacuum side.
The diode pins are connected through the PTFE to two SHV coaxial feedthroughs. Their  
air sides are directly connected to a custom-developed frontend electronics module.
It provides a high-voltage of $-1.0\,\mathrm{kV}$ to the diode to establish an almost radial electrical collection field between the grounded vessel and the diode, as depicted in Figure~\ref{fig:drift_sim}.
A battery installed in series provides the reverse bias voltage of $9\,\mathrm{V}$ to the diode.
The analog current signal of the diode is capacitatively decoupled from the high-voltage circuit and fed into a two-stage low-noise preamplifier with a total transimpedance gain of~$\sim 10^8\,\Omega$ and a bandwidth of~$100\,\mathrm{kHz}$.
Low- and high-pass filters 
reduce the electronic noise.
The shaped signals from the alpha decays of $^{218}\mathrm{Po}$ and $^{214}\mathrm{Po}$ with energies of $6.0\,\mathrm{MeV}$ and $7.7\,\mathrm{MeV}$ create amplitudes of $~500\,\mathrm{mV}$ and $~750\,\mathrm{mV}$, respectively.
The typical decay time of the signals is $\sim 50\,\mathrm{\upmu s}$.
They are digitized and analyzed by a 14-bit multichannel analyzer (CAEN DT5781a) sampling the signal at $100\,\mathrm{MS}/\mathrm{s}$.
An event is read out if the pulse exceeds a threshold set sufficiently low to be surpassed by any relevant alpha signal.
For every event, its timestamp, pulse height, and raw waveform data are stored.
Storage of the raw data could be disabled, however, especially during detector commissioning the direct access to the waveform data was very useful.

% SC and ELOG
During every measurement, which consists of the radon transfer and data acquisition phases, ambient and process parameters, such as temperatures and pressures, are monitored and stored in a database.  
A custom-developed lightweight slow control system running on an industry-grade microcontroller (KUNBUS RevPi Core 3$+$) is used for that purpose.

\section{Measurement Procedure}
\label{sec:detector_operation_and_model}

$^{226}\mathrm{Ra}$ has a long half-life of $1602\,\mathrm{years}$. $^{222}\mathrm{Rn}$ is thus assumed to be produced and emanated with constant radon emanation activity~$\mathcal{R}$.
The emanation rate $\mathcal{R}^{\mathrm{sample}}$ of a sample is determined by measuring the alpha decays of the radon daughter isotopes $^{214}\mathrm{Po}$ and $^{218}\mathrm{Po}$.
The standard measurement procedure with the radon emanation detector consists of three phases: \textit{radon emanation}, \textit{radon transfer}, and the \textit{polonium activity measurement}.
The activities of the various isotopes in these phases can be computed analytically by solving the respective rate equations;
the time evolution of every isotope depends on its radioactive decays and the decays of its mother isotopes.

% emanation
\paragraph{Radon Emanation:} The sample under study is closed off in the emanation vessel EV, which is subsequently evacuated and filled up with purified helium to atmospheric pressure.
Governed by the $^{222}\mathrm{Rn}$ half-life of $3.82\,\mathrm{d}$, the emanated radon activity within the emanation vessel asymptotically approaches the secular equilibrium activity $\mathcal{R}^{\mathrm{sample}}$.

% radon transfer
\paragraph{Radon Transfer:}
Typically after a few $^{222}\mathrm{Rn}$ half-lives, the accumulated radon atoms are concentrated and transferred from the emanation vessel EV into the detection vessel DV via the transfer trap TT.
%After a few $^{222}\mathrm{Rn}$ half-lives, the accumulated radon atoms are concentrated and transferred from the emanation into the detection vessel~DV via the transfer trap~TT. 
%Before the procedure, the transfer trap is cleaned from radon by repeatedly filling it with purified helium gas, followed by heating and evacuation. Then, the trap~TT is immersed in liquid nitrogen and the gas content of the emanation vessel evacuated through the trap, whereby the extracted radon atoms adsorb onto the porous charcoal. 
Prior to the procedure, the transfer trap is purged of radon by repeated filling with purified helium gas, followed by heating and evacuation. The transfer trap is then immersed in liquid nitrogen and continuously filled with about $1\,\mathrm{bara}$ of helium until the temperature and consequently the pressure in the trap stabilize. Finally, the gas content of the emanation vessel is evacuated through the transfer trap, and the extracted radon atoms are adsorbed onto the porous charcoal.
Afterwards, the trap is closed off, filled with purified helium, and then heated up to $175\,^{\circ}\mathrm{C}$ such that the radon atoms desorb and mix with the carrier gas.
By opening the line from the transfer trap to the previously evacuated detection vessel, the carrier gas and hence the majority of the radon atoms can expand into the detection vessel.
%By flushing purified helium through the TT, any remaining radon atoms are guided into the detection vessel until an absolute pressure of $1.0\,\mathrm{bara}$ is reached.
%
%The helium flow is currently neither monitored nor controlled.
Any remaining radon atoms are finally transported into the detection vessel by flushing purified helium through the transfer trap until an absolute pressure of $1.0\,\mathrm{bara}$ is reached. There is currently no direct measurement of the helium flow, but the pressure increase in the detection vesselis monitored and the flow is adjusted such that this last step takes roughly 2\,minutes.
%Instead, we only reduce the helium bottle pressure down to $1.6\,\mathrm{bara}$.
%Transferring the radon from the heated charcoal trap~TT into the detection vessel takes less than two minutes.

% polonium activity measurement
\paragraph{Polonium Activity Analysis:}
Most impurities, e.g., contaminants from outgassing and the radon decay products themselves, are removed by the hot getter during the transfer from the emanation vessel EV into the detection vessel DV.
Thus, the $^{218}\mathrm{Po}$ and $^{214}\mathrm{Po}$ sample signal activities inside the detection vessel, $A^{\mathrm{sample}}_{^{218}\mathrm{Po}}(t)$ and $A^{\mathrm{sample}}_{^{214}\mathrm{Po}}(t)$, are zero at the start of the measurement $t=t_{\mathrm{meas}}^{0}$, i.e., when the detection vessel is closed shut, 
%At the start of the measurement, they 
and increase until radon and polonium are in secular equilibrium.
%At the start of the measurement, $A^{\mathrm{sample}}_{^{214}\mathrm{Po}}(t)$ and $A^{\mathrm{sample}}_{^{218}\mathrm{Po}}(t)$ increase until radon and polonium are in secular equilibrium.
Once equilibrium is reached, the radon and polonium activities decrease according to the characteristic time scale of radon.
%Once the transfer is finished, the polonium sample activities increase due to the exponential radioactive decay of the trapped radon sample atoms. 
The $^{218}\mathrm{Po}$ and $^{214}\mathrm{Po}$ decays are identified by their respective energy, as shown in the right panel of Figure~\ref{fig:zeolite_granulate_measurement} on page~\pageref{fig:zeolite_granulate_measurement}. 
The number of detected %$^{218}\mathrm{Po}$ or $^{214}\mathrm{Po}$ 
events  $n^{\mathrm{meas}}$ from a certain polonium isotope during a measurement interval $\Delta t_{\mathrm{meas}}$ is given by
%\begin{align}
%    n^{\mathrm{meas}} &= \varepsilon^{\mathrm{det}} \, \int_{\Delta t_{\mathrm{meas}}}
%\hspace*{-3.0mm}A^{\mathrm{sample}}(t) 
%\,\mathrm{d}t \,\,+\bar{n}^{\mathrm{DV}}(\Delta t_{\mathrm{meas}}) \,\,+\bar{n}^{\mathrm{EV}}(\Delta t_{\mathrm{meas}})\quad , \label{eq:box_counting}
%\end{align}
%
\begin{align}
    n^{\mathrm{meas}} &= \bar{n}^{\mathrm{DV}}(\Delta t_{\mathrm{meas}}) \,\,+\bar{n}^{\mathrm{EV}}(\Delta t_{\mathrm{meas}}) \,\,+ \varepsilon^{\mathrm{det}} \, \int_{\Delta t_{\mathrm{meas}}}
\hspace*{-3.0mm}A^{\mathrm{sample}}(t;\mathcal{R}^{\mathrm{sample}}) 
\,\mathrm{d}t \quad , \label{eq:box_counting}
\end{align}
%
%\begin{align}
%    n^{\mathrm{meas}} &= \varepsilon^{\mathrm{det}} \, \int_{\Delta t_{\mathrm{meas}}}
%\hspace*{-3.0mm}A^{\mathrm{sample}}(t) 
%\,\mathrm{d}t \,\,+\bar{n}^{\mathrm{DV}} \,\,+\bar{n}^{\mathrm{EV}}\quad , \label{eq:box_counting}
%\end{align}
%\begin{align}
%    n^{\mathrm{meas}} &= \varepsilon^{\mathrm{det}} \, \int_{\Delta t_{\mathrm{meas}}}
%\hspace*{-3.0mm}[A^{\mathrm{sample}}(t)\,+A^{\mathrm{DV}}(t)\,+A^{\mathrm{EV}}(t) 
%]\,\mathrm{d}t \,\,\,\quad , \label{eq:box_counting}
%\end{align}
which takes into account the detection efficiency $\varepsilon^{\mathrm{det}}$ (see Section~\ref{ssec:calibration}) and the mean number of expected background events from both the detection vessel $\bar{n}^{\mathrm{DV}}$ and the emanation vessel $\bar{n}^{\mathrm{EV}}$  (see Section~\ref{ssec:background}). Since the polonium signal activities $A^{\mathrm{sample}}(t)$ can be expressed in analytical form (as the solution of a system of coupled, inhomogeneous first-order differential equations), Equation~\eqref{eq:box_counting} can be solved for the radon activity at $t_{\mathrm{meas}}^{0}$.
By additionally taking into account the duration of the radon emanation and transfer phases, one can then extrapolate the radon emanation activity $\mathcal{R}^{\mathrm{sample}}$ of the sample under study.

%%%%%%%%%%%%%%%%%%%%%%%%%%%%%%%%%%%
%%% Detector Performance
%%%%%%%%%%%%%%%%%%%%%%%%%%%%%%%%%%%

\section{Detector Performance}
\label{sec:performance}

In this Section, we present results on the performance of the radon emanation detector obtained during detector commissioning.

\subsection{Backgrounds}
\label{ssec:background}

During a sample measurement, both the detection vessel DV and emanation vessel EV also emanate $^{222}\mathrm{Rn}$ atoms, that, along with leakage of other decays into the region of interest and detector artifacts, contribute to the overall number of measured events $n^{\mathrm{meas}}$, as expressed by Equation~\eqref{eq:box_counting}.

% DETECTION VESSEL BACKGROUND:
%To measure the detection vessel background, it was filled with $1\,\mathrm{bara}$ of purified helium.
%Several such blank measurements were conducted.
%After approximately four weeks, average equilibrium detection vessel background rates of 2.4~counts per day (cpd; $28\,\mathrm{\upmu Hz}$) and 1.5~cpd ($17\,\mathrm{\upmu Hz}$) were measured in the regions of interest of the $^{218}$Po and $^{214}$Po channels, respectively.
%These rates comprise the sum of intrinsic $^{222}$Rn emanation from the detection vessel and additional components that potentially result from the leakage of other decays, such as those from the $^{220}$Rn chain, as well as detector-specific artifacts.
%Due to the a priori unknown time dependence of all these effects, the detection vessel background is determined in a data-driven manner for each sample measurement.
%We estimate the number of events, $\bar{n}^{\mathrm{DV}}(\Delta t_{\mathrm{meas}})$, that is attributed not to a sample but to the detection vessel, by averaging the number of background events that occurred during the equivalent measurement duration $\Delta t_{\mathrm{meas}}$ in the individual background measurements.
The detection vessel background was measured multiple times after being filled with $1\,\mathrm{bara}$ of purified helium. After approximately four weeks, average equilibrium background rates of 2.4~counts per day (cpd; $28\,\mathrm{\upmu Hz}$) and 1.5~cpd ($17\,\mathrm{\upmu Hz}$) were recorded in the regions of interest for the $^{218}$Po and $^{214}$Po channels, respectively.
These rates comprise the sum of intrinsic $^{222}$Rn emanation from the detection vessel and additional components that potentially result from the leakage of other decays, such as those from the $^{220}$Rn chain, as well as detector-specific artifacts. Because the background rates may be time-dependent, we calculate the expected number of background counts for each sample measurement separately. To do this, we determine the average number of events in the background-only measurements, $\bar{n}^{\mathrm{DV}}(\Delta t_{\mathrm{meas}})$, which were detected within the same window $\Delta t_{\mathrm{meas}}$ used for a given sample measurement. This number is subtracted from the number of sample events recorded.

% emanation vessel background
The radon emanation background rate of the emanation vessel $\mathcal{R}^{\mathrm{EV}}$ is determined in a measurement without sample, following the procedure outlined in Section~\ref{sec:detector_operation_and_model} and accounting for the detection vessel background determined above.
Two such background measurements were conducted.
For both measurements, both polonium channels are in excellent agreement with one another.
Taking into account the detection efficiencies $\varepsilon ^{\mathrm{det}}$ from Equation~\eqref{eq:detection_efficiency} yields a value of
\begin{align}
    \mathcal{R}^{\mathrm{EV}} = \left( 0.16 \pm 0.03 \right) \,\mathrm{mBq} \quad . \label{eq:ev_background}
\end{align}
%With this measured EV background activity $\mathcal{R}^{\mathrm{EV}}$, 
The expected number of emanation vessel background events is then computed by integrating the emanation vessel polonium activity model $A^{\mathrm{EV}}(t)$:
\begin{align}
\bar{n}^{\mathrm{EV}}(\Delta t_{\mathrm{meas}})=\int_{\Delta t_{\mathrm{meas}}} A^{\mathrm{EV}}(t; \mathcal{R}^{\mathrm{EV}})\,\mathrm{d}t \quad. \label{eq:ev_background_expectation}
\end{align}
We correct the time-dependent activity models for the emanation and transfer durations, which allows us to account for unavoidable temporal variations of the processes, caused by, e.g., different sample and emanation vessel sizes.
The $^{220}\mathrm{Rn}$ emanating from the sample and emanation vessels has a half-life of $55\,\mathrm{s}$ and is thus expected to decay during the sample transfer. %Any of its non-noble  progenies plate out onto the getter or the charcoal surfaces. $^{220}\mathrm{Rn}$-induced backgrounds from the emanation vessel or the sample itself are thus neglected.
All of its non-noble progenies will plate out on the getter or porous charcoal. $^{220}\mathrm{Rn}$-induced backgrounds from the emanation vessel or the sample itself are thus neglected.

\subsection{Detection Efficiency}
\label{ssec:calibration}

% measurement description
The overall detection efficiency $\varepsilon^{\mathrm{det}}$ is obtained by comparing the experimentally measured emanation rate with the known reference value of a calibrated sample.
The sample was provided by G.~Zuzel (Jagiellonian University, Krak\'{o}w, Poland).
It consists of two 2\,mm thick stainless steel discs with a diameter of 20\,mm, onto which $^{226}\mathrm{Ra}$ ions were electrodeposited.
The reference measurements of the same source were carried out by H.~Simgen at the Max-Planck-Institut für Kernphysik (MPIK) in Heidelberg, Germany, utilizing miniaturized proportional counters~\cite{gzhs09} and finding an emanation activity of $(47.6\pm 1.5)\,\mathrm{mBq}$.
The source is stored in a CF40 vacuum vessel, which can be closed off by two VCR bellow valves. For the calibration measurements, it was connected to the gas system, replacing the emanation vessel.

\begin{figure}[t!]
    \centering
    \includegraphics[width=\textwidth]{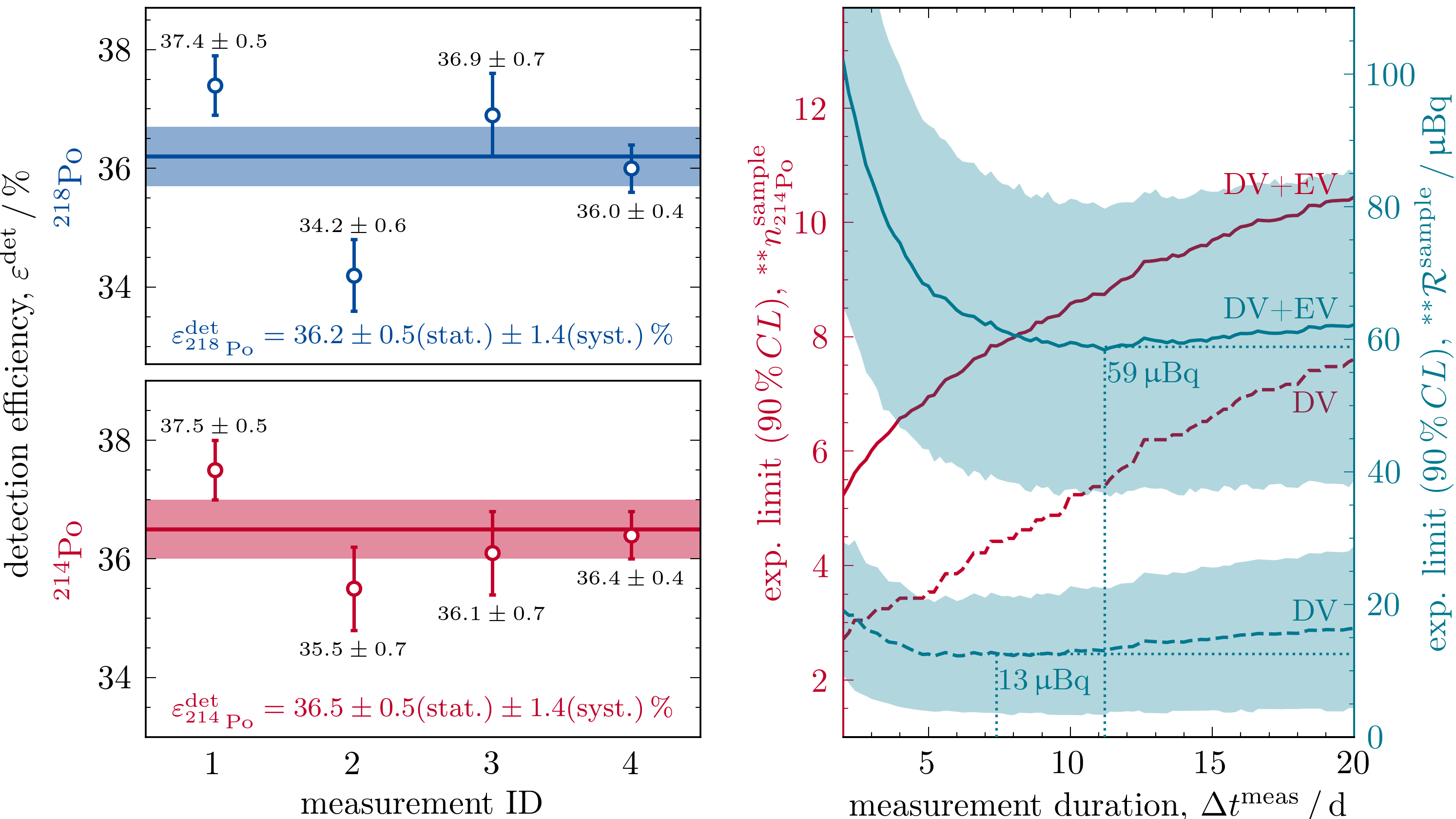}
    \caption{
        \textbf{(Left)} The results of four independent measurements of a $^{226}$Ra calibration source are shown relative to the absolute $^{222}$Rn emanation rate obtained from miniaturized proportional counters.
        The detection efficiency $\varepsilon^{\mathrm{det}}$, counting both $^{218}\mathrm{Po}$ (blue) and $^{214}\mathrm{Po}$ (red) decays, agrees in all four measurements, which also validates the radon concentration and transfer procedure.
        \textbf{(Right)} Estimation of MonXe's sensitivity in terms of the Poissonian box-counting model described in the text, considering only the $^{214}\mathrm{Po}$ background of the detection vessel (DV, dashed lines), and the backgrounds of the detection and the emanation vessels (DV+EV, solid lines). 
        In the absence of a signal, the expected median upper limits at $90\,\%\,\mathrm{C.L.}$ on the number of detected $^{214}\mathrm{Po}$ signal events ${^{***}n}^{\mathrm{sample}}_{^{214}\mathrm{Po}}$ (red lines, left axis) increase with increasing measurement time $\Delta t^{\mathrm{meas}}$.
        The corresponding upper limits at $90\,\%\,\mathrm{C.L.}$ on the radon emanation activity ${^{***}\mathcal{R}}^{\mathrm{sample}}$ of the sample (petrol lines, right axis) show optima at $13\,\mathrm{\upmu Bq}$ and $59\,\mathrm{\upmu Bq}$, respectively. (Upper limits are given as the median of the underlying Monte Carlo distribution; the shaded areas indicate their widths) .
    }
    \label{fig:detector_performance}
\end{figure}

% evaluation and discussion
The calibration campaign consisted of four individual measurements of the radon emanation rate of the calibrated sample, following the routine presented in Section~\ref{sec:detector_operation_and_model}.
The left panel of Figure~\ref{fig:detector_performance} shows the individual results relative to the reference value.
The measurement's statistical uncertainties result from the Poissonian process of counting the number of polonium decays.
%Apart from the second measurement (most visible in the $^{218}\mathrm{Po}$ channel), the data agrees within the uncertainties, supporting the hypothesis of constant efficiency.
We attribute the larger spread of the data to systematic variations in the manual transfer procedure, and adjust for this overdispersion by scaling the combined uncertainty by roughly a factor of two, such that the fit of a constant to all measurements yields $\chi_{\mathrm{red}}^2=1$.
Combining the consistent results of both polonium channels (the individual values are given in the left panel of Figure~\ref{fig:detector_performance}) we infer MonXe's detection efficiency to be
\begin{align}
  \mathcal{\varepsilon}^{\mathrm{det}} = \left( 36.3 \pm 0.2 (\mathrm{stat.}) \pm 1.4 (\mathrm{syst.})\right)\,\% \quad .
  \label{eq:detection_efficiency}
    %\mathcal{\varepsilon}^{\mathrm{det}}_{^{214}\mathrm{Po}} = 36.2 \pm 0.5 (\mathrm{stat.}) \pm 1.4 (\mathrm{syst.}) \,\% \quad\,\,\text{and}\quad\,\,  \mathcal{\varepsilon}^{\mathrm{det}}_{^{218}\mathrm{Po}} = 36.5 \pm 0.5 (\mathrm{stat.}) \pm 1.4 (\mathrm{syst.}) \,\, . 
%    \label{eq:detection_efficiency}
\end{align}
The systematic uncertainty is dominated by the $\sim 3\,\%$ uncertainty of the reference measurement.
Note that $\mathcal{\varepsilon}^{\mathrm{det}}$ already includes efficiency losses of at least $50\,\%$ due to the finite solid angle coverage of the active photodiode for alpha particles emitted on its surface.
The particle tracking simulation studies depicted in Figure~\ref{fig:drift_sim} indicate an electrostatic collection efficiency close to $100\,\%$ for the realized hemispherical detection vessel geometry and a static collection field generated by a high-voltage of $-1.0\,\mathrm{kV}$~\cite{wb21}.
We furthermore expect a transfer efficiency of roughly $100\,\%$.
By expanding and flushing the much smaller volume of the calibration source's emanation vessel directly into the detection vessel, we found the same radon activity as with the established transfer protocol. Also, small variations in the transfer temperature and helium flow do not alter the transfer efficiency.
A potential source of efficiency losses might be insensitive areas of the silicon PIN photodiode.
Nevertheless, the detection efficiency measured here suggests the fraction of initially recoil-ionized $^{218}\mathrm{Po}$ ions in helium being at least $72\,\%$, exceeding the value of $59\,\%$ found in earlier works~\cite{Howard1991}.

\subsection{Sensitivity}
\label{ssec:sensitivity}

% Poissonian counting model
We evaluate the detector sensitivity in terms of a single-bin Poisson counting experiment, following the analysis scheme outlined in Section~\ref{sec:detector_operation_and_model}:
The integer number of measured events $n^{\mathrm{meas}}$ is Poisson-distributed with the expected value $\bar{n}^{\mathrm{sample}}+\bar{n}^{\mathrm{DV}} +\bar{n}^{\mathrm{EV}}$.
Whether or not $n^{\mathrm{meas}}$ significantly exceeds the expected background $\bar{n}^{\mathrm{DV}}+\bar{n}^{\mathrm{EV}}$, given a significance level $\alpha = 0.1$, is determined by computing the corresponding $p$-value assuming the background-only hypothesis.
If $p<\alpha$, the background-only hypothesis is considered rejected by the data, and one can determine the sample's radon emanation rate from the excess number of signal events.

% setting an upper limit
If no signal above the background is observed, i.e., if $p\ge\alpha$, we quote the \textit{observed upper limit} $^{*}n^{\mathrm{sample}}$ on the number of signal events %at confidence level 
as the largest value of $\bar{n}^{\mathrm{sample}}$ that still yields less than $n^{\mathrm{meas}}$ detected events  with probability~$\alpha$:
\begin{align}
    p = \sum_{n=0}^{n^{\mathrm{meas}}} \mathrm{Poisson}(n\,;\, \bar{n}^{\mathrm{DV}}+\bar{n}^{\mathrm{EV}} + {^{*}}n^{\mathrm{sample}})  &\stackrel{!}{=} \alpha \quad .   
\end{align}
The observed upper limit is thus
\begin{align}
    {^{*}n}^{\mathrm{sample}} &= \frac{1}{2}\,F_{\chi^2}^{-1}\left[1-\alpha; 2(n^{\mathrm{meas}}+1)\right] -(\bar{n}^{\mathrm{DV}}+\bar{n}^{\mathrm{EV}}) \quad . \label{eq:observed_upper_limit}
\end{align}
In Equation~\eqref{eq:observed_upper_limit}, the sum of Poissonian probabilities is identified with the cumulative chi-squared distribution $F_{\chi^2}$, %(related via the definition of the Gamma function), 
which then allows computing ${^{*}\bar{n}}^{\mathrm{sample}}$ in analytical form.

% expected sensitivity
To estimate the experimental sensitivity, we compute the distribution of %a priori  
\textit{expected upper limits} on the detected number of signal events ${^{**}\bar{n}}^{\mathrm{sample}}$ at $90\,\%\,\mathrm{C.L.}$.
Monte Carlo data resembling the distribution of the number of observed events under the assumption of the background-only hypothesis is generated according to Equation~\eqref{eq:observed_upper_limit}.
The right panel of Figure \ref{fig:detector_performance} shows the median expected upper limit on the number of signal events from the sample ${^{**}\bar{n}}^{\mathrm{sample}}$ for the $^{214}\mathrm{Po}$ line and two different background contributions vs.~the measurement time.
Via the activity model~\eqref{eq:box_counting}, assuming infinite emanation time (which is approximately the case after an emanation period of four weeks) and infinitesimal transfer time, and taking into account the detection efficiency of Equation~\eqref{eq:detection_efficiency}, one can translate the expected upper limit of signal events into the corresponding 
%idealized 
upper limit on the radon emanation rate ${^{**}\mathcal{R}}^{\mathrm{sample}}$.
%The solid and dashed lines depict the cases where both DV and EV backgrounds or the DV background only are considered.

% discussion
The fluctuations in the right panel of Figure~\ref{fig:detector_performance}  are %not caused by too little statistics of the Monte Carlo data but result from 
due to the quantized number of expected detector vessel background events $\bar{n}^{\mathrm{DV}}(\Delta t_{\mathrm{meas}})$. %, which depends on the integer number of DV background events detected during $\Delta t_{\mathrm{meas}}$.
Initially, the ${^{**}\mathcal{R}}^{\mathrm{sample}}$ sensitivity curves steeply decrease until they reach a local minimum. For longer measurement times, the mean and width of the distributions show a steady increase.
This (at first glance counter-intuitive) time-dependence is caused by the asymptotically falling ratio of accumulated signal events (emanated by the sample and transferred once into the detection vessel) and background events (constantly emanated from the detection vessel walls):
While the signal activity will decrease exponentially with increasing measurement time as the sample decays, the background activity remains constant, once secular equilibrium is reached.
%while the signal events will initially dominate over the slowly and linearly increasing number of DV background events, they decrease exponentially and eventually fall below the number of accumulated DV background events. 
For the standard measurement procedure, i.e., a sample placed inside the emanation vessel EV, we hence quote the sensitivity of the MonXe radon emanation detector %to low sample emanation activities 
as the minimum of the curve taking into account the background from both vessels (EV+DV):
\begin{align}{^{**}\mathcal{R}}^{\mathrm{sample}} = 59 \, \mathrm{\upmu Bq} \quad . \label{eq:sensitivity}
\end{align}
As a consequence of the time behavior shown in Figure~\ref{fig:detector_performance}, a measurement is terminated, and an upper limit on the emanation rate of the sample is placed if no signal is detected after twelve days.
The theoretical detection vessel background-only sensitivity of 
\begin{align}{^{**}\mathcal{R}}^{\mathrm{sample}} = 13 \, \mathrm{\upmu Bq} \quad . \label{eq:sensitivity}
\end{align}
is only valid for radon emanation measurements of components such as vacuum vessels, getters, or valves that can be directly connected to the MonXe system.
For comparison, %today's most sensitive radon emanation instruments, 
the miniaturized proportional counters at the Max-Planck-Institut f\"ur Kernphysik (MPIK) in Heidelberg, Germany, which are among the most sensitive radon emanation instruments to date, achieve sensitivities of $\sim$40\,$\mathrm{\upmu Bq}$~\cite{gzhs09} with some counters exhibiting an intrinsic emanation activity up to four times lower than the detection vessel background rate of our detector.
The background emanation activity of the MPIK  $80\,\mathrm{l}$ emanation vessel is $(0.16\pm 0.05)\,\mathrm{mBq}$, which is similar to that of the four times smaller MonXe emanation vessel given in Equation~\eqref{eq:ev_background}.
The R.E.S.~facility at the South Dakota School of Mines and Technology uses a 
detector concept similar to MonXe and features two 
large emanation chambers of 13\,l and 300\,l. 
Its sensitivity of $200\, \mathrm{\upmu Bq}$~\cite{mb19} is comparable to 
the one of our instrument, 
while our vessel backgrounds and detection efficiency are slightly superior~\cite{lzcol20}.

%%%%%%%%%%%%%%%%%%%%%%%%%%%%%%%%%%%
%%% Sample Screening Measurements
%%%%%%%%%%%%%%%%%%%%%%%%%%%%%%%%%%%

\section{Sample Screening Measurements}
\label{sec:sample_screening_measurements}

In this section, we demonstrate the performance of the MonXe radon emanation detector based on two samples with very different emanation rates.

\subsection{High-Activity Sample: Zeolite Granulate}
\label{ssec:zeolite}

\begin{figure}
    \centering
    \includegraphics[width=\textwidth]{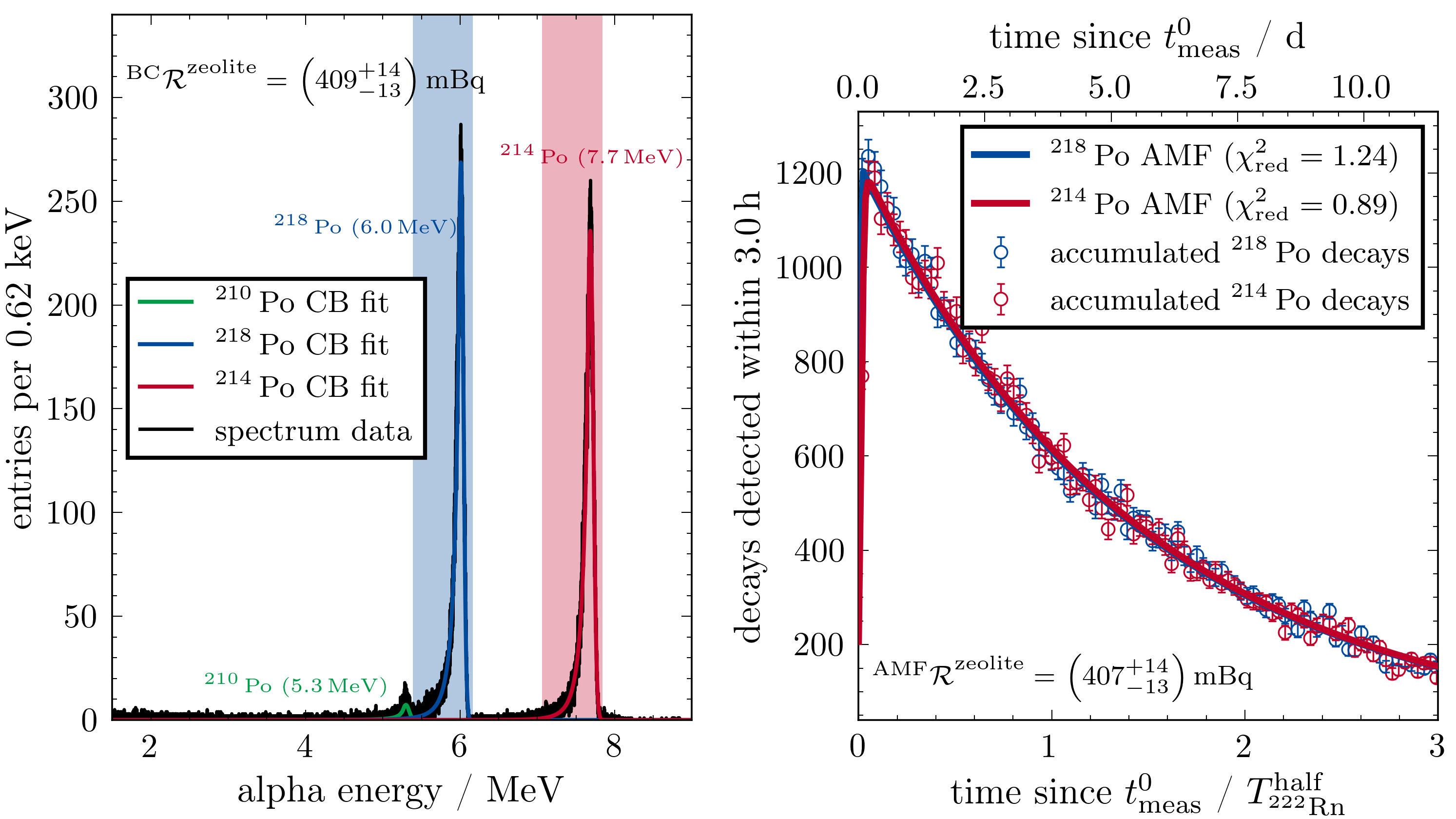}
    \caption{
        Radon emanation measurement of $730\,\mathrm{g}$ of commercial zeolite adsorbent pellets for 11.59\,days, following an emanation phase of 7.73\,days.
        \textbf{(Left)} The measured alpha spectrum %and standard box-counting (BC) analysis. 
        shows the two expected peaks from $^{218}\mathrm{Po}$ (blue) and $^{214}\mathrm{Po}$ (red); the $^{210}\mathrm{Po}$ peak (green) is strongly suppressed because of its much longer half-life. The peaks can be well described by Crystal Ball functions. The tail towards lower energies is caused by 
        angle-of-incidence-dependent energy losses in the PIN diode's insensitive p-layer. The colored areas define the energy windows for the event selection. 
        The number of $^{222}$Rn atoms at the start of the measurement can be computed analytically using the number of events observed in the peaks and the modeled polonium activity (box-counting analysis, BC).
%        Utilizing an analytic model of the polonium activities, the box-counting analysis yields a radon emanation activity of ${^{\mathrm{BC}}\mathcal{R}}^{\mathrm{ema}}_{^{222}\mathrm{Rn}} = \left( 409 _{-13} ^{+14} \right)\,\mathrm{mBq}$.
        \textbf{(Right)} Detected events in the energy regions of interest accumulated within $3\,\mathrm{h}$ intervals vs.~time. %Alternative activity model fit (AMF) analysis for high-activity samples.
        The activity model is fitted to the data of both isotopes (solid lines). The model describes the data very well and can also be used to obtain the initial number of $^{222}$Rn atoms (activity model fit, AMF). Both analysis methods yield identical results.
        %Accounting for the overall detection efficiency, one can extrapolate the zeolite sample's radon emanation activity of ${^{\mathrm{BC}}\mathcal{R}}^{\mathrm{zeolite}} = 410 \pm 4 (\mathrm{stat.}) \pm 16 (\mathrm{syst.})\,\mathrm{mBq}$.
    }
    \label{fig:zeolite_granulate_measurement}
\end{figure}

% measurement description
Zeolites are a class of microporous minerals typically used as an adsorbent, e.g., in backing pump adsorption traps to prevent the backstreaming of oil vapor.
The examined sample consists of $730\,\mathrm{g}$ of commercial zeolite adsorbent pellets (Pfeiffer Vacuum Technology AG Zeolith PK 001 248-T).
The left panel of Figure~\ref{fig:zeolite_granulate_measurement} shows the alpha energy spectrum acquired over a measurement period of $\Delta t_{\mathrm{meas}}= 11.59\,\mathrm{d}$ and after an emanation time of $7.73\,\mathrm{d}$.
One can clearly distinguish three peaks corresponding to the energies of the alpha particles emitted by $^{210}\mathrm{Po}$, $^{218}\mathrm{Po}$, and $^{214}\mathrm{Po}$, at $5.3\,\mathrm{MeV}$, $6.0\,\mathrm{MeV}$, and $7.7\,\mathrm{MeV}$, respectively. 
The long half-life of $^{210}\mathrm{Pb}$ of $22.3\,\mathrm{y}$ strongly suppresses the $^{210}\mathrm{Po}$ peak. The peak shapes of $^{218}\mathrm{Po}$ and $^{214}\mathrm{Po}$ are well described by a Crystal Ball function~\cite{cb_function_definition}.
Their low-energy tails are attributed to angle-of-incidence-dependent energy losses when the alpha particles traverse the insensitive p-layer of the PIN photodiode.
From the Crystal Ball fits, one can extract the energy resolution of~$1.5\,\%$, determined as the full width at half maximum of the $^{214}\mathrm{Po}$ peak.
The energy scale is defined by identifying the peak's mean values from the fit with the respective alpha energies.
The $^{218}\mathrm{Po}$ and $^{214}\mathrm{Po}$ events are selected from a predefined energy interval around the peak means.
These were chosen to include over $99\,\%$ of the Crystal Ball integrals while excluding the stray $^{220}\mathrm{Rn}$ events mentioned in Section \ref{ssec:background}.
Since the integral coverage proved to be robust against the fit uncertainties, the data were not corrected for the $<1\,\%$ selection inefficiency.
% box counting results
Following the box-counting (BC) analysis presented in Section~\ref{sec:detector_operation_and_model}, which takes into account the background contributions of the detection and emanation vessels, the specific radon emanation activity of the zeolite sample is 
\begin{align}
    {\mathcal{R}}^{\mathrm{zeolite}} = \left( 562 \pm 5(\mathrm{stat.}) \pm 22 (\mathrm{syst.})\right)\,\mathrm{\frac{mBq}{kg}} \quad . \label{eq:zeolite_bc_result}
\end{align}

% activity model fit
The box-counting analysis essentially ignores the knowledge of the individual event trigger timestamps.
However, for high-activity samples, this timing information can be utilized to validate the underlying model assumptions.
Instead of just counting all events recorded during the entire measurement $\Delta t_{\mathrm{meas}}$, one can subdivide $\Delta t_{\mathrm{meas}}$ in equal intervals and count the detected polonium events in each bin for $3\,\mathrm{h}$ intervals, as shown in the right panel of Figure~\ref{fig:zeolite_granulate_measurement}.
For each time interval, Equation~\eqref{eq:box_counting} applies, and the number of radon atoms at $t_{\mathrm{meas}}^0$ can be determined from the fit of the activity model to the data.
The reduced $\chi^2$-values of $1.24$ and $0.89$ indicate an excellent agreement between data and model for both $^{218}\mathrm{Po}$ and $^{214}\mathrm{Po}$.
Both analysis methods yield identical radon emanation rates.
%activities at the start of the measurement~$t_{\mathrm{meas}}^{0}$: $A^{\mathrm{zeolite}}(t_{\mathrm{meas}}^{0}) = \left(112.1 \pm 0.4\right) \, \mathrm{mBq}$ (box-counting) and $A^{\mathrm{zeolite}}(t_{\mathrm{meas}}^{0})=(112.2 \pm 0.5)\,\mathrm{mBq}$ (fit of model).
%For the box-counting result, the quoted uncertainty resembles the propagated Poisson uncertainty, whereas for the model fit analysis, the statistical and systematic uncertainties arising from the fit were combined in quadrature.
%The excellent agreement of data and model validates the detector model assumption, e.g., that the initially zero polonium population at $t_{\mathrm{meas}}^{0}$ only emerges from the decaying radon sample and the DV background.

% HPGe measurement
The bulk activity of the zeolite granulate was additionally measured with the high-purity germanium gamma-spectrometer GeMSE~\cite{dgrea22}.
The $^{226}\mathrm{Ra}$ activity was inferred from the $609.3\,\mathrm{keV}$ and $1765\,\mathrm{keV}$ gamma lines of $^{214}\mathrm{Bi}$ and assuming secular equilibrium among $^{226}\mathrm{Ra}$ and its short-lived daughters.
%To protect the procreated radon against the instrument's gaseous nitrogen purge, the sample was retained in gas-tight plastic bags.
To avoid the radon emanated from the sample being removed from the sample cavity by the gamma spectrometer's nitrogen purge, the sample was kept in gas-tight plastic bags.
%located at the Vue des Alpes underground laboratory~\cite{VdA} under a rock overburden of 620\,m water equivalent.
The measured $^{226}\mathrm{Ra}$ activity of $(11.4_{-0.7}^{+1.0})\,\mathrm{Bq}/\mathrm{kg}$ reveals that (at normal temperature and helium pressure, and assuming secular equilibrium) only $\sim$5\,\% of the  $^{222}\mathrm{Rn}$ atoms produced in $^{226}\mathrm{Ra}$ decays are emanated from the porous zeolite granulate.
%\begin{align}
    %{^{\mathrm{MonXe}}\mathcal{R}}^{\mathrm{zeolite}} = 0.562 \pm 0.005 (\mathrm{stat.}) \pm 0.022 (\mathrm{syst.})\,\mathrm{\frac{mBq}{kg}} \,\,\stackrel{\times 20}{<}\,\, 11.4_{-0.7}^{+1.0}\,\mathrm{\frac{Bq}{kg}} = {^{\mathrm{GeMSE}}A}_{^{226}\mathrm{Ra}} \,. \label{eq:comparison_monxe_gemse}
%\end{align}
%By grinding the porous material to finer particulates, one might increase the fraction of emanated radon atoms to unity.If realizable, this could allow configuring calibration sources with tunable emanation activity.

\subsection{Low-Activity Sample: Semi-Finished PTFE}
\label{ssec:semi_finished_ptfe}

% introduction: measuring PTFE
Polytetrafluoroethylene (PTFE, Teflon\textsuperscript{\tiny\textregistered}) is a widely used construction material in most low-background experiments due to its unique electrical insulation and optical properties.
PTFE semi-finished products are compression molded and sintered from granulated PTFE resin.
Here we present measurements of the radon emanation of two semi-finished PTFE samples: a sample manufactured by ElringKlinger~AG and a sample  manufactured by Fluorseals~S.p.A.. Each sample consisted of three cubic blocks (320\,$\times$\,160\,$\times$\,60\,mm$^3$ of $\sim$6\,kg mass each).

% measurement
In preparation for the measurements, a few microns were milled off from all surfaces of the blocks.
Because PTFE is expected to emanate only trace amounts of radon 
%Since PTFE is expected only to emanate trace amounts of radon 
\cite{XENON:2020fbs}, a series of $26$~narrow grooves of $20\,\mathrm{mm}$ depth were additionally saw-milled into the two largest faces of the cuboids to increase the total surface area from $0.48\,\mathrm{m}^2$ to $2.47\,\mathrm{m}^2$ per sample.
The blocks were cleaned in a bath of $6\,\mathrm{mol/l}$ nitric acid, then immersed in deionized water and ethanol, and finally blow-dried with pressurized helium.
To reduce potential outgassing of ambient radon that might have diffused into the porous material, the emanation vessel housing the cleaned PTFE samples was continuously evacuated with the turbomolecular pump for two weeks prior to the start of the actual emanation process of roughly three weeks.
By the start of the measurement, any contribution from ambient radon would have decreased by more than $95\,\%$.
Because of the additional evacuation, we assume the remaining fraction to be even smaller.
%, followed by the transfer, measurement, and analysis procedure described in Section~\ref{sec:detector_operation_and_model}.

% results
For both sample measurements, the $^{214}\mathrm{Po}$ and $^{218}\mathrm{Po}$ channels led to compatible results; the following radon emanation activities were measured:
\begin{align}
    \mathcal{R}^{\text{ElringKlinger}} =\left(61 _{-19}^{+18}\right) \,\mathrm{\frac{\upmu Bq}{m^2}} \quad\text{and}\quad 
    \mathcal{R}^{\text{Fluorseals}} = \left(34 _{-15}^{+14}\right) \,\mathrm{\frac{\upmu Bq}{m^2}} \quad . \label{eq:monxe_ptfe_measurements}
\end{align}
The systematic uncertainties from the calibration are a factor of ten smaller than the statistical uncertainties and are thus omitted in the following.
A second measurement of the Fluorseals sample %with the lower activity only 
 yielded upper limits of $51\,\mathrm{\upmu Bq}/\mathrm{m^2}$ and $43\,\mathrm{\upmu Bq}/\mathrm{m^2}$ for the $^{218}\mathrm{Po}$ and $^{214}\mathrm{Po}$ channels, respectively,  in agreement with the first measurement.
The Fluorseals detection corresponds to a total sample emanation activity of $(84 _{-37}^{+35})\,\mathrm{\upmu Bq}$ and thus lies only slightly above the theoretical optimum sensitivity of the detector of $59 \, \mathrm{\upmu Bq}$, given in Equation~\eqref{eq:sensitivity}.
Statistical fluctuations can easily move the measured activity above or below the detector's significance limit.

For reference, the PTFE reflectors used in the XENON1T dark matter experiment exhibit an emanation activity of $(24 \pm 5)$\,$\mathrm{\upmu Bq}/\mathrm{m}^2$~\cite{XENON:2020fbs}, which is comparable to the value inferred for the Fluorseals sample. 
These measurements were conducted with the miniaturized proportional counter infrastructure at the Max-Planck-Institut für Kernphysik in Heidelberg; a sample with a total mass of $32\,\mathrm{kg}$ and a surface area of $4\,\mathrm{m}^2$ was examined. 
The XENON1T reflectors were treated with a new diamond milling head to achieve a smooth surface to optimize the material's light reflectivity~\cite{XENON:2017lvq}.
Since radon emanation strongly depends on the surface properties of a sample, it is possible that this particular treatment led to an improved micro-porosity of the surface compared to the one of our saw-milled grooves.
With an emanation activity of $(12 {+5 \atop -10})$\,$\mathrm{\upmu Bq}/\mathrm{m}^2$, the PTFE used in the LZ dark matter experiment is also cleaner.
A very large sample of it was measured with the R.E.S.~facility mentioned above~\cite{lzcol20}.

%%%%%%%%%%%%%%%%%%%%%%%%%%%%%%%%%%%
%%% Conclusio
%%%%%%%%%%%%%%%%%%%%%%%%%%%%%%%%%%%

\section{Conclusion}
\label{sec:conclusio}

% motivation
The background of many rare-event search experiments is affected by the radioactive decays of $^{222}\mathrm{Rn}$ and its daughters.
As radon emanates from any detector construction material, quantifying the emanation rate of potential materials is crucial for optimizing the background of the next generation of low-background experiments.
%As radon is emanated off any detector construction material, quantifying the emanation rate of potential materials is crucial to optimize the background of next-generation of low-background experiments.
In this work, we present the design and performance of the MonXe radon emanation detector, designed to contribute to the radiopurity assay programs of the future astroparticle physics observatories DARWIN~\cite{darwin_16} and XLZD~\cite{Aalbers:2022dzr}.
%In this work, we presented the design and performance of the MonXe radon emanation detector, which will contribute to the radiopurity assay program of the future DARWIN~\cite{darwin_16} and XLZD~\cite{Aalbers:2022dzr} astroparticle physics observatories.

% MonXe summary
MonXe's detection concept is based on the spectrometric measurement of alpha decays of polonium atoms, which have been electrostatically collected on the surface of a silicon PIN photodiode.
By utilizing cryogenic physisorption traps, the radon atoms emanating from a  sample are transferred into a separate detection vessel. % in order to reproducibly screen samples of various sizes, shapes, and materials.
The alpha decays of the $^{222}\mathrm{Rn}$ daughters $^{218}\mathrm{Po}$ and $^{214}\mathrm{Po}$ are measured with an energy resolution of $1.5\,\%$ and an detection efficiency of about $\sim36\,\%$ per isotope. The sensitivity of the instrument, taking into account the measured backgrounds in the emanation and detection vessels, was determined as $59\,\mathrm{\upmu Bq}$ at $90\,\%$ C.L.. 
The performance of the MonXe radon emanation detector was demonstrated by determining the radon emanation of high-activity commercial zeolite granulate and two samples of semi-finished PTFE, with emanation rates close to the instrument's sensitivity.
% calibration

% outlook
At the time of writing, the MonXe detector is operated manually.
However, it is foreseen to automate the radon sample transfer from the emanation into the detection vessel by controlling new pneumatic valves and mass flow controller by the slow control system. A second independent emanation vessel is currently under construction and will decrease turnover times between measurements. The second vessel will be electropolished to possibly improve its intrinsic emanation, which currently limits the instrument's sensitivity.

%%%%%%%%%%%%%%%%%%%%%%%%%%%%%%%%%%%
%%% Acknowledgements
%%%%%%%%%%%%%%%%%%%%%%%%%%%%%%%%%%%

\section*{Acknowledgments}
This work was supported by the European Research Council (ERC) grant No.~724320 (ULTIMATE).
We gratefully acknowledge H.~Simgen and J.~Westermann (Max-Planck-Institut für Kernphysik, Heidelberg, Germany) for performing the reference calibration measurements, and G.~Zuzel (Jagiellonian University, Krak\'{o}w, Poland) for providing the calibration source.
We also thank the teams of the mechanical and electronics workshops of Institute of Physics, Freiburg, and in particular R.~Mori for the development of the preamplifier.
Finally, we thank all the Bachelor students and interns who contributed to commissioning the detector: J.~Alt,  W.~Boemke, L.~God, R.~Kirsch, and V.~Lieb.

%%%%%%%%%%%%%%%%%%%%%%%%%%%%%%%%%%%
%%% Bibliography
%%%%%%%%%%%%%%%%%%%%%%%%%%%%%%%%%%%

\bibliography{main.bib}{}
\bibliographystyle{JHEP}

\end{document}